\documentclass[12pt]{article}
\usepackage{cite,epsfig,amssymb,amsmath,graphicx,color}
\newcommand{\RNum}[1]{\uppercase\expandafter{\romannumeral #1\relax}}

\def\beq{\begin{equation}}
\def\eeq{\end{equation}}

\def\de{\delta}

\newcommand{\be}{\begin{equation}}
\newcommand{\ee}{\end{equation}}
\newcommand{\bear}{\begin{eqnarray}}
\newcommand{\eear}{\end{eqnarray}}
\newcommand{\ba}{\begin{array}}
\newcommand{\ea}{\end{array}}

\usepackage[normalem]{ulem}
\usepackage{varioref}

\begin{document}

\begin{center}
\large\bf Canonical  Quantization of Noncompact Spin System
\end{center}

\begin{center}
{\large Phillial Oh\\[3mm]

{\it Department of Physics and Institute of Basic Science,\\
        Sungkyunkwan University, Suwon 16419, Korea\\[2mm]

{\tt ploh@skku.edu}}}

\end{center}
\hspace{\fill}
\vspace{5mm}
\begin{flushleft}
\bf{Abstract}
\end{flushleft}

We consider spin system defined on the coadjoint orbit with noncompact
symmetry and  investigate the quantization. Classical  spin with noncompact  $SU(N,1)$ symmetry is first formulated as a dynamical system  and 
 the constraint analysis is performed to reduce the system from the group space to the coadjoint orbit which is a symplectic manifold with Kahler structure. We achieve this by solving the constraint directly.   It is shown that the dynamical variables describing the noncompact spins can be written as functions of canonically conjugate variables and canonical quantization is possible on the reduced phase space.  
With the quantum mechanical Hamiltonian acting on the holomorphic coherent state in Hilbert space, we obtain the exact propagator by  solving the time-dependent Schr\"odinger equation.

\vspace{4mm}

\noindent

\newpage
\section{Introduction}

One area application of noncompact symmetry is to the generalized spin with noncompact symmetry \cite{Kazakov:2018qez}. Noncompact spin variables are  defined as  generating functions of the noncompact symmetry in analogy with the compact spin. They can be given as a Lie-algebra valued functions of the  group element $g\in G$; Starting from a fixed Lie algebra-valued $x$, which belongs to the tangent space of the group manifold, one can consider  a group action $gxg^{-1}$ on $x,$ which 
 represent the adjoint transformation of the group $G.$ The Poisson bracket structure between the group variables can be introduced on the group space and using this, one can show that noncompact spins are generators of the symmetry. Reduction \cite{Guillemin:1990ew} from the   flat tangent space to the reduced phase space can be performed by treating the system as a constrained system, and   noncompact spins become dynamical variables on $G/H$ where $H$ is the stabilizer of the fixed element $x$. The coset space $G/H$
is the coadjoint orbit associated with $x$ and inherits a natural symplectic structure carried from the flat group space. 
Therefore, the coset space $G/H$  can be considered as a generalized phase space. Different choices of $x$ give different reduced phase spaces and symplectic structures. The quantum mechanics on the reduced  phase space gives unitary representations of the group $G$ as the quantum Hilbert space, and each symplectic structure corresponds to different unitary irreducible representation of the group $G$ upon quantization \cite{Woodhouse:1992de}.
%

In this paper, we consider noncompact spin in the case of
$SU(N,1)~(N\geq 2)$ symmetry and its quantization. The case for $SU(1,1)$ spin received much attention  mainly based on coherent state approach
\cite{Zhang:1990fy}, but explicit extension to higher rank group has been scarce.
Our main purpose will be to investigate  whether   canonical quantization, which is  usually  bypassed in favor of geometric quantization
 \cite{Woodhouse:1992de}\footnote{Also,   method of using oscillator representations of constrained dynamics  is another route to quantization \cite{Gates:1989im}.
} of  the generalized phase space, is possible. If available, it can make the quantization procedure more straightforward. We first write down a classical dynamical system  of  noncompact spin by presenting an explicit Lagrangian with $SU(N,1)$ symmetry from which Poisson structure can be obtained.  Treating the system with constrained dynamics, 1st and 2nd class constraints are identified. At this stage, two options are available. The first is to go through Dirac's constraint analysis to find the Dirac bracket. The quantization can be followed by replacing the Dirac brackets with Dirac commutators. Despite its widespread use, in our case it does not directly identify the local coordinate on the reduced phase space  with which geometrical quantities such as metric and noncompact spin can be expressed.
 Instead, we solve the constraints explicitly and go to the reduced phase space with physical degree of freedom only. For each 1st class constraint, which corresponds to gauge symmetry, one extra gauge fixing condition can be assumed in solving the constraint. 2nd class constraint has no such freedom.

After this process, the theory reduces to the coadjoint orbit 
\cite{Oh:1995av} of $SU(N,1)$ group.
In the case of $SU(2,1)$ it is either  
   complex projective space
$CP(1,1)=SU(2,1)/U(2)$ or flag manifold $SU(2,1)/U(1)\times U(1)$ depending on the stabilizer group $H$. Both of them are Kahler manifold with symplectic structure. All the geometrical quantities can be expresses as functions of the local complex coordinates $(\xi, \bar\xi)$ explicitly. There exist a well-defined geometric quantization  process on these manifold \cite{Woodhouse:1992de}. We investigate whether more conventional canonical quantization approach is viable.  The Poisson bracket can be introduced with the help of the inverse Kahler metric; $\{\bar\xi, \xi\}\sim g^{-1}$. Inspection of the Poisson bracket suggests existence of `conversion factor' $C(\xi, \bar\xi)$  which yields  canonically conjugate momentum, when it multiplies the coordinates; $\{C\bar\xi, \xi\}\sim 1, \{\bar\xi, C\xi\}\sim 1. $ Therefore one  can define $P\equiv C\xi$ and $\bar P\equiv C\bar\xi$ as canonical conjugate momenta of $\bar\xi$ and $\xi$.  The conversion factor itself has 
well-defined Poisson bracket relations with these canonically conjugate pairs and can equivalently be replaced by these variables. 
As far as phase space is concerned, one has freedom to  choose  ${\cal P=(\bar P, \xi)}$ or ${\cal \bar P}=(P, \bar\xi)$  as canonical phase space. Once a choice is made,  it is shown that the dynamical variables describing the noncompact spin can be written as function  of canonically conjugate variables on the phase space. And the classical dynamics   defined on the generalized phase space can be transformed into canonical one. When  quantization is considered,  it  has to be first decided  which phase space to work with.
Then,  one must select polarization of the physical state.  It is shown that after phase space and polarization are chosen, that is, holomorphic or antiholomorphic, the expression of the noncompact spin in terms of   canonically conjugate  variables is unique except the possible normal ordering problem. Adopting a prescription  for normal ordering, 
 canonical quantization can be pursued on the reduced phase space by replacing the Poisson bracket with Dirac commutator; $\{~,~\}\rightarrow i[~,~]$ .  

Then, we consider canonical quantization with the coherent states. The coherent state for  the noncompact $SU(1,1)$
was discussed in a large number of papers \cite{inomata1992}, 
but the extension to  higher rank $SU(N,1)$
was nontrivial that hampered such attempts \cite{Girelli:2017dbk}.  
Conventional  coherent state $\vert \xi>$ with unit norm can be explicitly constructed, but it carries a normalization factor $N(\xi,\bar\xi)$
which depends on both $\xi$ and $\bar\xi$ with it. It could be a source of over-specification problem in quantization. Therefore, we  consider holomorphic coherent state without the normalization factor. Both correspond to the discrete series of unitary irreducible representation of $SU(2,1)$ \cite{Bars:1989bb}.
We construct the quantum mechanical Hamiltonian operators obtained through canonical quantization which act on the coherent states in the Hilbert space and set up the time-dependent Schrodinger equation. 
 It is shown  that the  time-dependent Schrodinger equation can be
solved exactly for the Hamiltonian which corresponds to torus action on the reduced phase space $CP(1,1)$.
 
 The paper is organized as follows:
  In Sec. (2), explicit reduction to the coadjoint orbit of noncompact group is carried out through constraint analysis. In Sec. (3), we identify canonically conjugate variables and canonical quantization is done on the reduced phase space with the holomorphic coherent state. We also calculate the exact quantum mechanical propagator for an exactly soluble Hamiltonian. Sec (4) contains summary and discussions.

 .

\section{Reduced Phase Space of Noncompact Spin}

We start a brief summary of noncompact spin on coadjoint orbits.
They are coset space $G/H,$ where the group $G$ is the symmetry group and $H$ is the stabilizer of each point of the orbit.
For the group $G=SU(N,1)$, the coadjoint orbits    
can be classified into  
${\cal O}_{\{n_1,n_2,\cdots,n_l\}}
\equiv SU(N,1)/SU(n_1)\times\cdots\times SU(n_l)\times U(1)^{l}$ \cite{Duval:1981js}.
Here  $\sum_{i=1}^ln_i=N$  is equal to the rank of the group.
 Each corresponds to the orbit generated from a   reference  point $x$ in the Lie algebra
 as 
\begin{equation}
X=g^{-1}xg,~~g\in SU(N,1),\label{point}
\end{equation}
where
\begin{equation}
x=i\mbox{diag}(\underbrace{x_{1},\cdots, x_{1}}_{n_1},
\underbrace{x_{2},\cdots ,x_{2}}_{n_2},
\cdots,\underbrace{x_{l},\cdots, x_{l}}_{n_l},x_{N+1})~~(x_i\neq x_j)\label{fixedel}
\end{equation}
is a fixed antihermitian  traceless matrix with  $x_{N+1}=-\sum_{i=1}^{l}n_ix_i$.
$X$ is an arbitrary point of the coadjoint orbit ${\cal O}_{\{n_1,n_2,\cdots,n_l\}}$. We assume $x_1> x_2> \dots > x_l$ without loss of generality. 
When $n_1=N $ or 
$n_1=1$ and $x_2=\cdots=x_l=x_{N+1}$, 
the maximum-stability subgroup is $SU(N)\times U(1),$ and the orbit
corresponds to the minimal orbit which is a noncompact complex projective space
 $CP(N-1,1)=SU(N,1)/SU(N)\times U(1)$.
When $n_1= n_2= \cdots =n_l=1$, it corresponds to the maximal orbit
which is a 
noncompact flag manifold, $SU(N,1)/U(1)^N$. Taking the exterior derivative of the equation of Eq. (\ref{point}), we have
\begin{equation}
dX=[X,\theta],
\end{equation}
where the one-form $\theta=g^{-1}dg$ with values in the Lie algebra defines a canonical  one-form $\omega$  by
\beq
\omega={\rm Tr}( x\theta)={\rm Tr}(xg^{-1}dg). \label{canonicalone}
\eeq
Symplectic structure on the coadjoint orbit is inherited from  $\Omega=d\omega$
of (\ref{canonicalone}), which is a closed
and non-degenerate two-form. It is invariant under the left action of the group $SU(N,1)$ and right action of the stabilizer subgroup $H$ . 

The generator  $T_a$ of the $SU(N,1)$ group satisfies  
$[T_a, T_b]=if_{abc}T^c,$ and they are given by
\beq
T_a=\frac{\lambda_a}{2} ~~~(a=1,2,\cdots, N^2-1, N^2+2N ),
\nonumber
\eeq
if the generators belong to the subgroup $SU(N)\times U(1).$
The rest is organized as follows:
\begin{eqnarray}
T_{N^2}&=&\frac{i}{2}\lambda_{N^2+1}, ~~~T_{N^2+1}=-\frac{i}{2}\lambda_{N^2}, \\
T_{N^2+2}&=&\frac{i}{2}\lambda_{N^2+3}, ~~~T_{N^2+3}=-\frac{i}{2}
\lambda_{N^2+2},\\
& \cdots&\\
T_{N^2+2N-2}&=&\frac{i}{2}\lambda_{N^2+2N-1}, ~~~T_{N^2+2N-1}=-
\frac{i}{2}\lambda_{N^2+2N-2},
\label{gend}
\end{eqnarray}
where $\lambda_a$ is the generalized Gell-Mann matrices of $SU(N+1)$ group.
This  rearrangement
of the coset generators has the advantage that the commutation relations of the generators in terms of shifting operators have a simple relation with the compact $SU(N+1)$ algebra.  We normalize ${\rm Tr} (T_aT_b)=1/2\eta_{ab}$ where the raising and lowering of index $a, b, \cdots$ are performed with the metric:
\begin{equation}
\eta_{ab}=\mbox{diag}(\underbrace{{1},\cdots, {1}}_{N^2-1},
\underbrace{-1,\cdots ,-1}_{2N},
1).
\end{equation}
 That is, it is $+1$ when the indices $a,b$
belong to the subgroup $SU(N)\times U(1)$, and $-1$ otherwise.

They also inherit  the complex structure  from
the complex representation
 of ${\cal O}_{\{n_1,n_2,\cdots,n_l\}}
 =SL(N+1, {\bf C})/P_{\{n_1,n_2,\cdots, n_l\}}$ \cite{Picken:1988fw}.
Here,   $SL(N+1, {\bf C})$ is the complexification
of $SU(N,1)$ and $P_{\{n_1,n_2,\cdots, n_l\}}$ is a parabolic subgroup
of  $SL(N, {\bf C})$ which is the subgroup
of block upper triangular matrices in the
$(n_1+n_2+\cdots +n_l)\times (n_1+n_2+\cdots +n_l)$ block decomposition.
  Together with the symplectic structure, the coadjoint orbits  
${\cal O}_{\{n_1,n_2,\cdots,n_l\}}$ become the K\"ahler manifolds
with the symplectic two form  given
in the local complex coordinate $( \xi_\alpha,\bar \xi_{\bar\beta})$
by the Kahler form
\begin{equation}
\Omega=d\omega=i\sum_{\alpha,\beta}g_{\alpha\beta}d \xi^\alpha\wedge d\bar \xi^{\beta}.\label{sympe}
\end{equation}
In this section, we will explicitly carry out reduction from the Lie algebra-valued $X$ of (\ref{point}) to the
local complex coordinate $( \xi_\alpha,\bar \xi_{\beta})$
through constraint analysis. 

The metric $g_{\alpha\beta}$ can be expressed in terms
of K\"ahler potential $K$ by
\begin{equation}
g_{\alpha\beta}=\frac{\partial}
{\partial \xi^\alpha}\frac{\partial}{\partial\bar \xi^{\beta}}K.\label{nonfubini}
\end{equation}
Then, the Poisson bracket can be defined via
\begin{equation}
\{f,g\}=i\sum_{\alpha,\beta}g^{\alpha\beta}\left(\frac{\partial
f}{\partial \bar \xi^\alpha}\frac{\partial g}{\partial \xi^{\beta}}-
\frac{\partial g}{\partial\bar \xi^{\alpha}}
\frac{\partial f}{\partial \xi^{\beta}}\right).
\label{pbracket}
\end{equation}
Noncompact spin on the coadjoint orbit is defined as 
\begin{equation}
Q=igx g^{-1}\equiv 2 Q^a T_a ,\label{nspin}
\end{equation}
where $Q^a$ is the spin component.
$Q^a$ can be expressed as a function of the complex coordinate $( \xi_\alpha,\bar \xi_{\beta})$ after reduction and it can be shown that they  realize the $SU(N,1)$ algebra
upon using the Poisson bracket (\ref{pbracket}):
\begin{equation}
\{Q^a, Q^b\}=-f^{ab}_{\ \ c}Q^c.\label{susym}
\end{equation} 
We will explicitly demonstrate this reduction of symmetry algebra in the case of $SU(2,1)$  case.


We present a detailed description of the reduction for $SU(2,1)$ but generalization to higher $N\geq3$ is immediate. Let us consider  the element $g$ of $SU(2,1)$
expressed as 
 \beq
g=\begin{pmatrix}
\alpha_1 & \beta_1 & \gamma_1 \\	
\alpha_2 & \beta_2 & \gamma_2 \\
\alpha_3 & \beta_3 & \gamma_3 	\end{pmatrix}
,\label{sumatrix}
\eeq
where $\alpha_i, \beta_i, \gamma_i$ are arbitrary complex numbers.
The reference point is given by
\beq
x=i{\rm diad} (x_1, x_2, x_3),~x_1+x_2+x_3=0.\eeq
The unitary condition  $g^\dagger m g=m$ with the metric $m={\rm diag}(-1, -1, 1)$
gives the following constraints
\begin{eqnarray}
\psi_1&=&\sum_i\alpha_i\bar\alpha^i-1=0, ~~\psi_2=\sum_i\beta_i\bar\beta^i-1=0, ~~\psi_3=\sum_i\gamma_i\bar\gamma^i-1=0, ~~\nonumber\\
\phi_1&=&\sum_i\alpha_i\bar\beta^i=0, ~~\phi_2=\sum_i\beta_i\bar\gamma^i=0, ~~\phi_3=\sum_i\gamma_i\bar\alpha^i=0. ~~\label{constraints}
\end{eqnarray} 
The raising and lowering of the component indices are done with the metric $m_{ij}$. 
Using (\ref{sumatrix}), the canonical one-form (\ref{canonicalone}) becomes
(assuming non-vanishing $x_i$)
\beq
 \omega_1=
ix_1\bar\alpha^id\alpha_i+ix_2
\bar\beta^id\beta_i +ix_3\bar\gamma^id\gamma_i.\label{canone}
\eeq
The noncompact spin of (\ref{nspin}) can be expressed as
\begin{equation}
Q=-x_1 \vert \alpha>< \alpha\vert 
-x_2\vert\beta><\beta\vert-x_3 \vert \gamma>< \gamma\vert,
\label{matrixiso}
\end{equation}
where $\vert \alpha>, \vert\beta>, $ and $\vert\gamma>$ are column vectors with components $\alpha_i, \beta_i$ and $\gamma_i$ respectively, and
 $<\alpha\vert, <\beta\vert,$
and $<\gamma\vert$ are row vectors with components $\bar\alpha^i, \bar\beta^i$ and
 $\bar\gamma^i: <\alpha\vert\alpha>=<\beta\vert\beta>=<\gamma\vert\gamma>=1, <\alpha\vert\beta>=
 <\beta\vert\gamma>=<\gamma\vert\alpha>=0.$
The noncompact spin component
  $Q^a$ of Eq. (\ref{nspin}) is given by
\begin{equation}
Q^a= -x_1 < \alpha\vert T^a\vert \alpha>
-x_2<\beta\vert T^a\vert\beta>-x_3 < \gamma\vert T^a\vert \gamma>.\label{isoff}
\end{equation}
Defining Poisson bracket from (\ref{canone}) by 
 \begin{equation}
\{\bar \alpha^i, \alpha_j\}=(i/x_{1})\delta^{i}_{j}, 
~\{\bar\beta^i, \beta_j\}=(i/x_2)\delta^{i}_{j},
~\{\bar \gamma^i, \gamma_j\}=(i/x_{3})\delta^{i}_{j},
\label{canonif}
\end{equation}
we find $Q^a$ satisfies (\ref{susym}) and therefore, definition (\ref{nspin}) qualifies. Note also that  ${\rm Tr}Q^2=2Q^aQ_a=2(x_1x_2+x_2
x_3+x_3x_1)>0$ and 
${\rm Tr}Q^3=-3x_1x_2x_3$, which are related with the two Casimir invariants of the group $SU(2,1)$ \cite{Bars:1989bb}. 

The first step in reduction is to use  the orthonormal properties of unitary matrix and eliminate $\gamma_i$ via $\gamma_i=\epsilon_{ijk}\bar\alpha^j \bar\beta^k$ with $\epsilon_{123}=1.$
One can check that ${\rm det}g=1$ using this $\gamma_i.$
Eliminating $\gamma_i$'s, we find  the one-form  (\ref{canone}) becomes 
\begin{equation}
\omega_2=
iJ_1{\big(}\bar\alpha^id\alpha_i-d\bar\alpha^i\alpha_i{\big)}
+iJ_2{\big(}\bar\beta^id\beta_i-d\bar\beta^i\beta_i{\big)},
\label{ss1}
\end{equation}
where $J_1=x_1+\frac{1}{2}x_2, J_2=x_2+\frac{1}{2}x_1.$
We define a dynamical system of noncompact spin with the Lagrangian
\begin{equation}
L=i\sum_{i=1}^3\Big[J_1\big(\bar\alpha^i\frac{d\alpha_i}{dt}-\frac
{d\bar\alpha^i}{dt}\alpha_i\big)
+J_2\big(\bar\beta^i\frac{d\beta_i}{dt}-\frac{d\bar\beta^i}{dt}\beta_i\big)
\Big]
-H(Q^a)+\sum_{s=1,2}\lambda_s\psi_s+(\eta\phi_1+{\rm h.c}),\label{lagg}
\end{equation}
using the canonical one-form  (\ref{ss1}).
 In most cases, 
the Hamiltonian of the dynamical system can be taken as 
\begin{equation}
H=\sum_{a, b}c_{ab}Q^{a}Q^b +\sum_a c_aQ^a.\label{hamt}
\end{equation}
$c_{ab}$ and $c_a$ could depend on time, in general.
We assume that  $c_{ab}$ is chosen so  that the quadratic part of the Hamiltonian is positive-definite. For example, $c_{ab}=\eta_{ab}$ gives $Q_aQ^a$ which is  greater than zero.
To perform the  constraint analysis, let us first consider the Poisson bracket relations from (\ref{ss1}) given by \begin{equation}
\{\bar \alpha^i, \alpha_j\}=(i/2J_{1})\delta^{i}_{j}, 
~\{\bar\beta^i, \beta_j\}=(i/2J_{2})\delta^{i}_{j},\label{canoni}
\end{equation}
with the constraints $\psi_1$, $\psi_2$ and $\phi_1$ of Eq. (\ref{constraints}).
 In order to construct the constraint algebra on the reduced phase space, one can resort to Dirac's method. 
Let us  suppose $x_1\neq x_2$.
Using Eq. (\ref{canoni}), we can check that the following
constraint algebra holds ($\phi_1\equiv\phi$):
\begin{eqnarray}
\{\psi_p,\psi_q\}&=& 0, \quad (p,q =1,2)\nonumber\\
\{\psi_1,\phi\}&=&\frac{i}{2J_1}\phi,
 ~\{\psi_3,\phi\}=-\frac{i}{2J_{2}}\phi,\nonumber\\
\{\phi,\bar\phi\}&\approx&-\frac{i}{2}
\big(\frac{1}{J_1}-\frac{1}{J_2}\big).\label{constr}
\end{eqnarray}
We see that each of $\psi_p$'s is a first class constraint.  $\phi$ and $\bar\phi$ are second class in the case of
 $x_1\neq x_2$, and first class when $x_1=x_2$.  So the dimension of the reduced phase space, which is equal to minus twice the number of first class constraints and minus the number of second class constraints is 6 (4) for $x_1\neq x_2~(x_1=x_2).$

In passing, we mention that one can easily generalize to the coadjoint orbit ${\cal O}_{\{n_1,n_2,\cdots,n_l\}}
=SU(N,1)/SU(n_1)\times\cdots \times SU(n_l)\times U(1)^{l}.$ First,  
the last row can be eliminated by using the orthonormality condition. Then $g\in SU(N,1)$  has $2N(N+1)$ real components and the total number of constrints are $N^2.$ Among these, the number of 1st class constraints is  
$ N+\sum_{i=1}^ln_i(n_i-1),$ where the second term is the sum of the constraints belonging to each block $n_i.$
Then,  the number of 2nd class constraints is $N(N-1)-\sum_{i=1}^ln_i(n_i-1),$
and we 
obtain the dimension of the reduced
phase space as $2N(N+1)-2[N+\sum_{i=1}^ln_i(n_i-1)]-
[N(N-1)-\sum_{i=1}^ln_i(n_i-1)]=N(N+2)-\sum_{i=1}^ln_i^2,$
 which coincides with
the dimension of the orbit ${\cal O}_{\{n_1,n_2,\cdots,n_l\}}
.$

Having identified the 1st and 2nd class constraints, one can proceed by using Dirac method to quantize the system. The process fulfils the SU(N,1) symmetry via 
Eq. (\ref{susym}), but it does not yield geometrical information such as metric and symplectic structute in terms of intrinsic coordinates on the phase space.
This can be achieved by solving the constraints directly, if possible and eliminate the redundant variables in (\ref{lagg}) 
\cite{Faddeev:1988qp}.  Therefore, we go directly to the reduced phase space by solving the constraints. 
We first introduce supplementary conditions in number equal to that of the first class constraints such that they yield  a non-degenerate matrix of all  Poisson brackets of  the constraints and the supplementary conditions. For $x_1\neq x_2,$ the conditions
\begin{equation}
\pi_1\equiv\alpha_3-\bar\alpha_3=0, ~ \pi_2\equiv\beta_1-\bar\beta_1=0
\end{equation}
serve the purpose with ${\rm Re} (\alpha_3)\neq 0$ and ${\rm Re}(\beta_1)\neq 0.$ Introducing 
\begin{equation}
\alpha_1=\xi_1\alpha_3, ~\alpha_2=\xi_2\alpha_3, ~\beta_2=
\eta_1\beta_1, ~ \beta_3=\eta_2\beta_1,\label{cc1}
\end{equation}
 we find 
\begin{equation}
\alpha_3=\frac{1}{\sqrt{1-\vert\xi_1\vert^2-\vert\xi_2\vert^2}}, ~~ 
\beta_1=
\frac{1}{\sqrt{\vert\eta_2\vert^2-\vert\eta_1\vert^2-1}},\label{cc2}
\end{equation} 
upon using $\bar\alpha^i\alpha_i=\bar\beta^i\beta_i=1.$ Then the remaining constraints $\phi=0$ can be easily solved as 
\begin{equation}
\eta_2=\bar\xi_1+\bar\xi_2\eta_1,\label{cc3}
\end{equation}
thus eliminating $\eta_2$. The reduced coordinates $\xi_1,\xi_2, \eta_1$ describe the six-dimensional noncompact flag manifold $SU(2,1)/U(1)\times U(1).$ The compact version $SU(3)/U(1)\times U(1)$ is
 described in Ref. \cite{Picken:1988fw} by using complex line bundle method.
When $x_1=x_2$, we can add two more constraints of the form
$\pi_3\equiv\beta_2-\bar\beta_2=0, ~\pi_4\equiv\beta_3-\bar\beta_3=0.$  
We can introduce the same coordinate $(\xi_1, \xi_2, \eta_1, \eta_2)$, and find that $\eta_1$ and $\eta_2$ can now be expressed as 
\begin{equation}
\eta_1=\frac{\xi_1-\bar\xi_1}{\bar\xi_2-\xi_2},~
 \eta_2=\frac{\xi_1\bar\xi_2-\bar\xi_1\xi_2}{\bar\xi_2-\xi_2},
\end{equation} 
which become real variables.
Therefore, the reduced phase space is four dimensional with two complex coordinate $\xi_1$ and $\xi_2$, which is $CP(1,1)$.
Note that the real $\beta_i's$ nullify the second terms in (\ref{ss1})
 and these variables are redundant on $CP(1,1)$. 

Let us calculate the symplectic  structure from (\ref{ss1}) on the reduced phase space. From here on $\bar\eta_1\equiv \xi_3. $
Substituting Eqs. (\ref{cc1})-(\ref{cc3}) into (\ref{ss1}), we obtain the canonical one-form as
\begin{eqnarray}
\omega&=&iJ_1\Big[\frac{\xi_1d\bar\xi_1+\xi_2d\bar\xi_2-{\rm h.c}}
{1-\vert\xi_1\vert^2-\vert\xi_2\vert^2}\Big]+
iJ_2\Big[\frac{-\xi_3d\bar\xi_3+(\xi_{1}+
\xi_2\xi_3)(d\bar\xi_1+d\bar\xi_2\bar\xi_3+\bar\xi_2d\bar\xi_3)-{\rm h.c}}
{\vert\xi_1+\xi_2\xi_3\vert^2-\vert\xi_3\vert^2-1}\Big]\nonumber\\
&\equiv& \frac{i}{2}(\bar\partial-\partial)W,
\end{eqnarray}
where $W$ is the Kahler potential given by
\begin{equation}
W=-2J_1\ln (1-\vert\xi_1\vert^2-\vert\xi_2\vert^2)+2J_2\ln
(\vert\xi_1+\xi_2\xi_3\vert^2-\vert\xi_3\vert^2-1).\label{kkkk}
\end{equation}
The symplectic two-form $\Omega=i\partial\bar\partial W $ of  $CP(1,1)$ is given by $(J_1\equiv J)$
\begin{equation}
\Omega=2iJ\Big[\frac{(1-\vert\xi_2\vert^2)d\xi_1\wedge d\bar\xi_1
+\bar\xi_1\xi_2d\xi_1\wedge d\bar\xi_2+\bar\xi_2\xi_1d\xi_2\wedge d\bar\xi_1
+(1-\vert\xi_1\vert^2) d\xi_2 d\wedge\bar\xi_2}
{(1-\vert\xi_1\vert^2-\vert\xi_2\vert^2)^2}\Big].\label{FSM}
\end{equation}
The symplectic two-form for the flag manifold can be also  written down explicitly.   
Note that these expressions of symplectic two-form have much resemblance with the compact case \cite{Picken:1988fw} except the characteristic  minus signs reflecting the noncompactness. As far as the concrete expression of the metric is concerned, the constraint analysis method yields the results without too much technical details. 


\section{Canonical quantization}

For dynamical analysis and quantization,  we focus on the $CP(1,1)$ manifold with $x_1=x_2 $ in  the fixed element $x$ of (\ref{fixedel}) which is written as 
$x=i \mbox{diag}(x_1,x_1,-2x_1)$.
Classical $SU(2,1)$ symmetry can be well described on
  complex projective space
$ CP(1,1)$ which is a
symplectic manifold and therefore could be considered to be the
phase space of classical mechanics. The Lagrangian of Eq. (\ref{lagg}) on the reduced phase space is given by 
\beq
L=i J\frac{{ \xi_{\alpha}}\dot{\bar \xi}_{\alpha}-{\bar \xi}_{\alpha}\dot \xi_\alpha}
{1-\vert\xi_1\vert^2-\vert\xi_2\vert^2}- H(Q).\label{actact}
\eeq
The noncompact spin of (\ref{matrixiso}) after elimination of  $\gamma_i$
 consists of contributions both from $\alpha_i$ and $\beta_i.$ 
However,  $\beta_i$ contribution does not fulfil  $SU(N,1)$ symmetry and  
it is an gauge artifact. Consequently, the noncompact spin component which realizes $SU(N,1)$ on $CP(1,1)$ 
symmetry is given by
\beq
Q^a=-2J< \alpha\vert T^a \vert \alpha>
=-\frac{2J}{1-\vert \xi_1\vert^2-\vert \xi_2\vert^2}
\left( T^a_{33}+T^a_{3\alpha}\xi_\alpha- \bar\xi_\beta T^a_{\beta3}
-\bar\xi_\alpha T^a_{\alpha\beta}\xi_\beta\right).\label{defii}
\eeq
where  the generators of the group $SU(2,1)$ of Eq. (\ref{gend})
are given 
\begin{eqnarray}
T_a=\frac{\lambda_a}{2} ~~~(a=1,2,3,8 ),~~
T_4=i\frac{\lambda_5}{2}, ~T_5=-i\frac{\lambda_4}{2},~T_6=i\frac{\lambda_7}{2},~
T_7=i\frac{\lambda_6}{2},\label{generates}
\end{eqnarray}
where $\lambda_a$ is the Gell-Mann matrices. They satisfy
\begin{eqnarray} 
[T_a, T_b]&=&if_{ab}^{~~c} ~T_c,\nonumber\\
\{T_a, T_b\}&=&\frac{1}{3}\eta_{ab}I+d_{ab}^{~~c}T_c.
\end{eqnarray}
$f_{ab}^{~~c}$ is the totally anti-symmetric 
structure constant and  $d_{ab}^{~~c}$
is totally antisymmetric \cite{Greiner:1989eu}. The raising and lowering of index $a, b, \cdots$ are done with
 the metric $\eta_{ab}={\rm diag}(1,1,1,-1,-1,-1,-1,1)$. With these generators, we have, for example\beq
Q^3=J\frac{\vert\xi_1\vert^2-\vert\xi_{2}\vert^2}{1-\vert \xi_1\vert^2-\vert \xi_2\vert^2}
,~~~~Q^8=\frac{J}{{\sqrt 3}}\frac{2+\vert\xi_1\vert^2+
\vert\xi_2\vert^2}{1-\vert \xi_1\vert^2-\vert \xi_2\vert^2}.\label{isodia}
\eeq

The Kahler structure of (\ref{kkkk}) in $CP(1,1)$ is given by 
\beq
W=-2J\ln(1-\vert\xi_1\vert^2-\vert\xi_2\vert^2). ~~~
\eeq
The metric (\ref{nonfubini}) from the  symplectic structure gives the noncompact version of Fibini-Study metric of (\ref{FSM})  by 
\beq
g_{\alpha\beta}=
2J\frac{(1-\vert\xi_1\vert^2-\vert\xi_2\vert^2)\de_{\alpha\beta}+
{\bar \xi}_{\alpha}\xi_{\beta}}{(1-\vert\xi_1\vert^2-\vert\xi_2\vert^2)^2},
\eeq
and the inverse $g^{\alpha\beta}$ satisfying $g_{\alpha\beta}
g^{\beta\gamma}=\de_{\alpha\gamma}$ is given by
\beq
g^{\alpha\beta}=\frac{1}{2J}(1-\vert\xi_1\vert^2-\vert\xi_2\vert^2)
(\de_{\alpha\beta}-{\bar \xi}_{\alpha}\xi_{\beta}).\label{metric}
\eeq
Using the above noncompact Fubini-Study metric, we    define the fundamental commutators from (\ref{pbracket}) as follows;
\beq
\{{\bar \xi}_{\alpha},\xi_{\beta}\}=\frac{i}{2J}(1-\vert \xi_1\vert^2-\vert \xi_2\vert^2)
(\de_{\alpha\beta}+{\bar \xi}_{\alpha}\xi_\beta),\label{poi3}
\eeq
\[\{\xi_\alpha,\xi_\beta\}=\{{\bar \xi}_{\alpha},{\bar \xi}_{\beta}\}=0.\]
The above Poisson bracket generates the following useful relations:
\beq
\{\bar\xi_\alpha, 
\frac{2J\xi_\beta}{1-\vert\xi\vert^2}\}
=\{\frac{2J\bar\xi_\alpha}{1-\vert\xi\vert^2}, \xi_\beta\}=i\delta_{\alpha\beta},
~\{\frac{2J\bar\xi_\alpha}{1-\vert\xi\vert^2}, 
\frac{\xi_\beta}{1-\vert\xi\vert^2}\}=\frac{i}{1-\vert\xi\vert^2}
(\delta_{\alpha\beta}+\bar\xi_\alpha\xi_\beta).\label{useful}
\eeq
Using these relations, we can  check that the noncompact spin 
$Q^a$ of (\ref{defii}) fulfils the $SU(N,1)$ symmetry of Eq. (\ref{susym}) after the reduction.
Hamiltonian vector field associated with each $Q^a$ of (\ref{defii}) is defined  by
\beq
X^a\rfloor \Omega+dQ^{a}=0.\label{moment}
\eeq
Using Eq. (\ref{defii}), we obtain
\beq
X^a=-i\left[(T^a)_{\alpha 3}+
(T^a)_{\alpha\beta}\xi_{\beta}
-(T^a)_{33}\xi_\alpha-(T^a)_{3\beta}\xi_\beta\xi_\alpha
\right]\frac{\partial}{\partial \xi_\alpha} + (c.c).
\eeq
It generates the following transitive action of the group $SU(2,1)$ on the $CP(1,1)$ manifold: 
\beq
\delta_{X^a}\xi_\alpha=\epsilon[(T^a)_{\alpha 3}+
(T^a)_{\alpha\beta}\xi_{\beta}
-(T^a)_{33}\xi_\alpha-(T^a)_{3\beta}\xi_\beta\xi_\alpha].
\eeq
Note that the action induces a linear transformation for generators belonging to the subgroup $SU(2)$, while the generators belonging to the coset
$G/H$ are being nonlinearly realized.

Eq. (\ref{useful}) provides essential information about canonical formulation. The first two relations   implies that 
\beq
\bar P_\alpha=\frac{2J\bar\xi_\alpha}{1-\vert\xi\vert^2},
\eeq
are canonical conjugate of the variables $\xi_\alpha$. And the same for$ P_\alpha$ to $\bar\xi_\alpha.$ 
Note that the Lagrangian (\ref{actact}) can be written in a canonical form like
\beq
L=\frac{i}{2}( P_{\alpha}\dot{\bar \xi}_{\alpha}-{\bar P}_{\alpha}\dot \xi_\alpha)
- H(Q),
\eeq
thus confirming $\bar P_\alpha (P_\alpha)$ as canonical conjugate of 
$\xi_\alpha (\bar\xi_\alpha).$
We can define canonical phase space including $P_\alpha$ or 
$\bar P_\alpha.$ However, all of them, 
$\xi_\alpha, \bar\xi_\alpha, P_\beta, \bar P_\beta$ cannot comprise
the phase space, because it double the dimension of the phase space. It is obvious that there are two options; One can choose canonical phase space either ${\cal P}=(\bar P, \xi)$ or ${\cal P}=(P, \bar\xi)$. 
After that,  the immediate obstacle is that $Q^a$ of (\ref{defii}) cannot be written as a function of  $(\bar P, \xi)$ or $(P, \bar\xi)$; for example, $Q^a$ contains the factor $C(\xi, \bar\xi)\equiv2
J(1-\vert\xi_1\vert^2-\vert\xi_2\vert^2)^{-1}.$  One way to avoid this difficulty is to neglect $\bar P_\alpha$ and $P_\alpha$ all together and resort to the geometric quantization  which quantize directly the reduced phase  described by local coordinates $(\xi, \bar\xi)$. The canonical quantization can be pursued, however, if one  notice that the  factor $C(\xi, \bar\xi)=2
J(1-\vert\xi_1\vert^2-\vert\xi_2\vert^2)^{-1}
$  has a well-defined Poisson bracket structure with $\xi_\alpha, \bar\xi_\alpha, P_\beta, \bar P_\beta$;
 \beq
\{\bar\xi_\alpha,C\}=i\bar\xi_\alpha, ~\{P_\alpha,C\}=-iP_\alpha;
~\{C, \xi_\beta\}=i\xi_\beta,~\{C, \bar P_\beta\}=-i\bar P_\beta\label{converion}.
\eeq
Therefore the conversion factor $C$ can be replaced by $\bar P_\alpha\xi_\alpha$ on ${\cal P}$
and $P_\alpha\bar\xi_\alpha$ on $\bar {\cal P}$~\footnote{Note that writing $C=P_\alpha/\xi_\alpha$ or $\bar P_\alpha/\bar\xi_\alpha$ does not help, because it mixes ${\cal P}$ and $\bar{\cal P}$ }.
With this recipe, it can be readily check that the noncompact spin (\ref{defii}) can be written as a function of canonical variables on each phase space uniquely.
In summary, 
we have argued that 
the classical dynamics on the generalized phase space described by $\xi$ and $\bar\xi$ can be also considered as the one with canonical phase space given either by
\beq
{\cal P}=(\bar P_\alpha, \xi_\alpha), ~ C=\bar P_\alpha\xi_\alpha;~~ 
\bar{\cal P}=( P_\alpha, \bar\xi_\alpha), ~~ C= P_\alpha\bar\xi_\alpha
\eeq

In quantizing the system, a choice whether to work with ${\cal P}$ or $\bar{\cal P}$ has to be made  first. Then,  upon replacing the Poisson bracket $\{f,g\}$ to Dirac bracket $i[f, g]$, $\bar P_\alpha$ or $P_\alpha$ can be replaced by
\beq
\bar P_\alpha\rightarrow \frac{\partial}{\partial\xi_\alpha},~~
P_\alpha\rightarrow -\frac{\partial}{\partial\bar\xi_\alpha}.
\eeq
The first replacement corresponds to holomorphic quantization, and the second to antiholomorphic. We are interested in the antiholomorphic quantization.
We observe that once a quantization scheme is selected, the quantum mechanical operator   $\hat Q^a$ corresponding to the classical function of  (\ref{defii}) can be  uniquely written as product of $\bar P_\alpha$ and   $\bar\xi_\alpha$
except it   depends on the normal ordering prescription one chooses. For example, when translating the classical function $P_\beta \bar \xi_\alpha$ into a quantum mechanical operator, we can have  
 either $\hat P_\beta\xi_\alpha$ or $\bar\xi_\alpha \hat P_\beta,$
 or many others as one chooses.
The only criteria available is that the they have to fulfill the symmetry algebra. 
 We choose to write in the antiholomorphic quantization, 
\beq
\hat Q^a=-
\left( T^a_{33}\bar\xi_\alpha \hat{P_\alpha}  +
T^a_{3\alpha}\hat P_\alpha-T^a_{\beta3} \bar\xi_\beta
\bar\xi_\alpha \hat{P_\alpha}
-T^a_{\alpha\beta}\bar\xi_\alpha\hat P_\beta \right),\label{defiii}
\eeq
where the prescription is chosen in such a way that momentum is arranged next  to the coordinates. Then, the quantum mechanical differential operators corresponding to the operators (\ref{isodia})
are given by
\beq
\hat Q^3=-\bar\xi_1\frac{\partial}{\partial \bar\xi_{1}} 
+\bar\xi_2\frac{\partial}{\partial \bar\xi_{2}} +J_{(3)},
 ~\hat Q^8=-{\sqrt 3}\bar\xi_1\frac{\partial}{\partial \bar\xi_{1}} 
-{\sqrt 3}\bar\xi_2\frac{\partial}{\partial \bar\xi_{2}}
+J_{(8)},\label{quntumiso}
\eeq 
where  $J_{(3)}, J_{(8)}$ which are constants associated with the normal ordering   are zero with this prescription. 
 Thus, the antiholomorphic  representation of the noncompact spin $\hat Q^a$ of  (\ref{defiii}) can be expressed as follows:
\begin{equation}
\hat Q^a(\bar\xi_\alpha)=[T^a_{3\alpha}-T^a_{\alpha\beta}\bar\xi_\alpha+
T^a_{33}\bar\xi_\alpha-T^a_{\beta 3}\bar\xi_{\beta}\bar\xi_\alpha]
\frac{\partial}{\partial\bar\xi_\alpha}.
\label{cpop}
\end{equation}
The above differential operator satisfy $[\hat Q^a(\xi),
\hat Q^b(\xi)]
=if^{ab}_{~~ c}\hat Q^c(\xi)$.

The time-dependent Schrodinger equation can be written as 
\beq
i\hbar\frac{\partial \Psi(\bar\xi)}{\partial t}=H(\hat Q)\Psi(\bar\xi), 
~~\Psi(\bar\xi)=<\xi\vert\Psi>.
\eeq 
As an application of our formalism, we calculate the propagator for a system which is exactly soluble. First, the state $\vert\xi>$ is chosen as a coherent state defined by\begin{eqnarray}
  |\xi \rangle= 
\exp \left[{\sqrt 2}\xi_1 E_{+2}+{\sqrt 2 }\xi_2E_{+3}\right]  |
\Lambda_0 \rangle.\label{aco}
\end{eqnarray}
Here,  the reference state  $|\Lambda_0 \rangle$  belongs to discrete series of the unitary irreducible representations of the $SU(2,1)$ group 
which  were investigated  in detail in Ref. \cite{Bars:1989bb}.   To have the  correspondence with the noncompact $CP(1,1)$ manifold, the state $\mid\Lambda_0\rangle $ must be annihilated by the maximal subgroup $SU(2) $
$(T_1, T_2, E_{\pm 1}) $ and the shifting operators $E_{-2}, E_{-3}$ defined by
\beq
E_{\pm 1}=\frac{T_1\pm iT_2}{\sqrt 2},~E_{\pm 2}
=\frac{T_4\pm iT_5}{\sqrt 2},~E_{\pm 3}=\frac{T_6\pm iT_7}{\sqrt 2}.
\eeq
It is also an eigenstate of the two commuting generators $H_1=T_1, H_2=T_8.$
Therefore, $\mid\Lambda_0\rangle $ 
is taken as the lowest spin state which is a $SU(2)$ singlet \cite{Bars:1989bb}
which has the following eigenvalues of $H_1$ and $H_2:$
\begin{equation}
H_1\mid\Lambda_0\rangle =0,
 ~~H_2\mid\Lambda_0\rangle =J \mid\Lambda_0\rangle.
\end{equation}
The coherent state (\ref{aco})  shows clearly the analytic properties  and have the advantage of being holomorphic in the parameter
$\xi$, which is useful for antiholomorphic quantization. But it  not normalized, and it  can be explicitly shown  \cite{poh} that the reproducing kernel of the coherent state is given by
\begin{equation}
<\bar \xi^\prime\vert\xi>=e^{W(\bar\xi^\prime,\xi)},
\label{normalization}
\end{equation}
where 
\beq
W(\bar\xi^\prime, \xi)=-2J\ln(1-
\bar\xi^\prime_1\xi_1-\bar\xi^\prime_2\xi_2).
\eeq
Consequently, the inner product 
$< \xi^\prime\vert\xi> $ is holomorphic functions of
the variable $\xi$ and anti-holomorphic functions of the variable $\bar\xi^\prime$.

We calculate the  propagator 
with the Hamiltonian (\ref{hamt}). We focus on exactly soluble case in which $c_{ab}=0$ and only $c_3$ and $c_8$ are non-vanishing with $Q^3$ and $Q^8$ given by Eq. (\ref{isodia}). Therefore, we consider 
\beq
H=\frac{\bar\omega_1\mid\xi_1\mid^2+\bar\omega_2\mid\xi_{2}\mid^2}{1-\vert \xi_1\vert^2-\vert \xi_2\vert^2},\label{hamttg}
\eeq
up to an irrelevant constant.
The equations of motion derived from the action (\ref{actact})
are those of two harmonic oscillators
although the Hamiltonian (\ref{hamttg})
appears to be highly nonlinear:
\begin{equation}
\dot{\bar\xi}_\alpha(t)-i\omega_\alpha\bar\xi_{\alpha}(t)=0,
\quad \dot\xi_\alpha(t)+i\omega_\alpha\xi_\alpha(t)=0; ~(\omega_\alpha
=\bar\omega_\alpha/J)\label{equation}
\end{equation}
The solutions are given by
\begin{equation}
\bar\xi_\alpha(t)=\bar\xi^{\prime}_\alpha e^{-i\omega_\alpha
(t-t^{\prime})},\quad
\xi_\alpha(t)=\xi_\alpha^\prime e^{i\omega_\alpha(t-t^\prime)}
\label{soll}
\end{equation}
Let us calculate the coherent state propagator. 
 Consider 
the propagator 
\begin{equation}
K(\bar\xi^{\prime},\xi;t)
= \langle \xi^{\prime}\vert e^{-i{\hat H}t}\vert
\xi\rangle.
\end{equation}
Note that the polarization is
chosen such that $K_2(\bar\xi^{\prime},\xi;t)$ is a
function  of $\bar\xi^{\prime}$ and  
$\xi$.
Hence  we get the following
time-dependent Schr\"odinger equation corresponding to the Hamiltonian (\ref{hamttg}): 
\begin{equation}
i\frac{\partial}{\partial t}
K(\bar\xi^{\prime},\xi;t)
=-\Big(\omega_1\bar\xi_1^{\prime}\frac{\partial}{\partial\bar\xi_1^{\prime}}
+\omega_2\bar\xi_2^{\prime}\frac{\partial}{\partial\bar\xi_2^{\prime}}\Big) K(\bar\xi^{\prime},\xi;t)~~~(t>0)
\end{equation} 
with  the boundary condition 
\begin{equation}
K_{2}(\bar\xi^{\prime},\xi;t){\Big\vert}
_{t\rightarrow 0}=\langle \xi^{\prime}\vert 
\xi\rangle=(1-\bar\xi^{\prime}_1
\xi_1-\bar\xi^{\prime}_2\xi_2)^{-2J}.
\end{equation}
One can check that the solution is given by
\begin{eqnarray}
K(\bar \xi^{\prime},\xi;t)
=\left( 1-\bar\xi^{\prime}_1\xi_1
e^{i\omega_1t}- \bar\xi^{\prime}_2\xi_2
e^{i\omega_2t}\right)^{-2J}.~~~\label{fapr}
\end{eqnarray}
The above expression has  resemblance to the compact case which  reproduces the Weyl character formula, 
 for example, in the $SU(2)$ case \cite{KeskiVakkuri:1991ta}
 and coherent state propagator for $SU(3)$ flag manifold \cite{Kim:1995wf}.
It would be interesting  to extend the analysis to the more  general case of nonvanishing time-dependent $c_{ab}(t) $ in  (\ref{hamt}) and investigate the exact solutions further.

\section{Summary and Discussion}

In summary, we formulated   dynamical noncompact spin system with 
$SU(2,1)$ symmetry and presented  canonical  quantization with the help of coherent states. The reduced phase space $CP(1,1)$ is  a nonlinear homogeneous space with transitive $SU(2,1)$ group action on it. On this space, we were able to identify canonically conjugate pairs of  dynamical
variables and the conversion factor which has well-defined representation in terms of  the canonically conjugate variables.
  It was shown that once polarization is chosen for quantization, the expression of the noncompact spin in terms of functions of these variables is unique except the normal ordering problem, and canonical quantization is viable. When the quantum mechanical Hamiltonian acting on the coherent state in Hilbert space is a linear combination of the generators associated with stabilizer of $CP(1,1)$, we were able to obtain an  exact propagator by  solving the time-dependent Schr\"odinger equation.

We note that the operator version of noncompact spin, for example, those of (\ref{cpop}) is usually obtained by the geometric quantization method. However, in our approach of canonical quantization, we have a general expression (\ref{defiii}) written as product
of canonically conjugate pairs if one replaces the conversion factor in terms of canonical variables. This factor is a  remnant of reduction; $\alpha_3$ in (\ref{canonif})
starts out as an independent variables, but  when it is eliminated through constraint, it  becomes a function on the phase space  that has a well-defined representation
in terms of the canonical variables. Then, the canonical quantization process can be pursed with the introduction of coherent state as Hilbert space. The procedure like prequantization in geometric quantization can be bypassed. It could be conceived as a mere technical merit, but more importantly, in this approach  quantization on generalized phase space is more or less straightforward just like the conventional canonical quantization. 

Our approach shares somewhat same spirit with Ref. \cite{kusnezov1989} which also studies canonical quantization of generalized phase space  in a different method.  It can be immediately inferred that such feature persists on the compact $CP(N)$ manifold. In general Kahler manifold with  potential $W$, the canonical momenta corresponding to $\xi$and $\bar\xi$ are $\frac{\partial W}{\partial\xi}$ and $\frac{\partial W}{\partial\bar\xi}$, respectively, but it remains to be seen whether canonical quantization process advocated in this work  carries through.  
 
 We conclude with a final remark.
Our main interest was $CP(1,1)$ obtained with $x_1=x_2.$
Being such, it does not cover all the irreducible representation of the group $SU(2,1).$
When  $x_1$ and $x_2$ are not equal, 
the reduced phase space becomes noncompact flag manifold $SU(2,1)/T$, where 
$T=U(1)\times U(1)$ is the maximal torus in  $SU(3).$
 In this case,  noncompact version of the  well-known  Borel-Weil-Bott theorem
 \cite{Witten:1987ty} assures us that all the discerete series of  irreducible representation
 of $SU(2,1)$
can be associated with the reduced phase space equipped with a symplectic
structure. It would be interesting if the present analysis can be extended to the noncompact flag manifold.

\section{Acknowledgments\\ }
\vspace{-0.3cm}
This work was supported by Basic Science Research Program through
the National Research Foundation of Korea (NRF) funded by the
Ministry of Education (Grant No. 2015R1D1A1A01056572).

\end{document}